\documentclass[aps,prl,twocolumn,groupedaddress,showkeys]{revtex4}
\usepackage{epsfig}

\begin{document}


\title{Unfolding mechanism and the free energy landscape of a single stranded DNA i-motif}

\author{Jens Smiatek$^1$}
\author{Chun Chen$^2$}
\author{Dongsheng Liu$^2$}
\author{Andreas Heuer$^1$}
\affiliation{$^1$Institute of Physical Chemistry, University of Muenster, D-48149 M{\"u}nster, Germany\\
	     $^2$Department of Chemistry, Tsinghua University, Beijing 100190, P. R. China}


\begin{abstract}
We present Molecular Dynamics simulations of a single stranded unprotonated DNA i-motif in explicit solvent. Our results indicate that the native structure in non-acidic solution at 300 K is unstable and completely 
vanishes on a time scale up to 10 ns. 
Two unfolding mechanisms with decreasing connectivity between the initially interacting nucleobases can be identified where one pathway is characterized as entropically more favorable.
The entropic preference can be mainly explained by strong water ordering effects due to hydrogen bonds for several occurring structures along the pathways.
Finally we are able to indicate via free energy calculations the most stable configurations 
belonging to distinct hairpin structures in good agreement to experimental results. 
\end{abstract}

\date{\today}

\maketitle
\section{Introduction}
The appearance of non Watson-Crick like structures in DNA has been reported two decades ago \cite{Gueron1993}. Since this time a lot of effort has been spent to investigate
these conformations and possible applications in detail \cite{Leroy2000,Trent2008,Liu2009,Mergny2002}. Experiments lead to the conclusion that these structures are the only known DNA configurations that involve 
systematic base intercalation \cite{Leroy2000}. Prominent representatives are the G-quadruplex structures and the i-motif \cite{Mergny2002} 
where the first one is formed by 
guanine (G) rich sequences \cite{Trent2008} while the latter is present in more cytosine (C) rich strands of DNA \cite{Leroy2000}.\\ 
The stabilizing mechanism for these at a first glance fragile structures is realized by a proton mediated cytosine binding between different strands or regions 
of the sequence resulting in a stable C-CH$^+$ pairing \cite{Gueron1993,Leroy2000,Liu2009,Mergny2002}. Due to an acidic environment,  
this is achieved by hemi-protonated cytosines which mimick an ordinary C-G binding as it is present in double helix DNA. 
Hence it becomes clear that these structures are only occurring at slightly acidic to neutral conditions resulting
in pH values from 4.8 to 7.0 \cite{Gueron1993,Leroy2000,Tan2005}. I-motifs show a remarkable stability \cite{Tan2005} and have been found as tetrameric and dimeric complexes
although their existence has also been proven for single stranded DNA \cite{Leroy2000}. 
A sketch of the C-CH$^+$ complex where the additional proton mediates a hydrogen bond between the nitrogens of the cytosine groups and the corresponding single stranded i-motif with its 
sequence is shown in 
Fig.~\ref{fig1}.\\
Due to its biological appearance in centromeric and telomeric DNA, the distinct i-motif conformations have been discussed as a new class of possible biological targets for cancers 
and other diseases \cite{Gupta1997,Hurley2001}.
However, a detailed investigation of the function in the human cell is still missing.
Despite this lack of knowledge, the application of this special configuration in modern biotechnology has experienced an enormous growth over the last years \cite{Liu2009}.  
Since the i-motif becomes unstable at pH values larger than 7, a systematic decrease and increase of protons in the solution by changing the pH value results in a reversible folding and unfolding mechanism.
It has been shown that this process occurs on a timescale of seconds \cite{Liu2009,Tan2005}.\\
Technological applications for this mechanism are given by molecular nanomachines \cite{Liu2009,Liu2003}, switchable nanocontainers \cite{Mao2007}, pH sensors to detect the pH value inside living cells \cite{Krishnan2009}, 
building materials for logic gate devices \cite{Qu2010} and sensors for distinguishing single walled and multi-walled carbon nanotube systems \cite{Qu2009}.
Recently it has been reported \cite{Zhang2010}, that the grafting density massively influences the structure of an i-motif layer in nanodevices due to steric hindrance.
Regarding these examples it becomes clear that a detailed investigation of the unfolding pathway of the i-motif is of prior importance.\\ 
In this paper we present the results of Molecular Dynamics simulations concerning the unfolding mechanism of a maximum unstable single stranded DNA i-motif structure without hemi-protonated cytosines. 
Our results indicate a fast initial decay of 
the i-motif leading to hairpin structures on a timescale up to 10 ns which dominate the unfolded regime in contrast to a fully extended strand. The numerical findings are validated by experimental 
Circular Dichroism (CD) spectropolarimetry data. By distinct investigation of the unfolding pathways, two main mechanisms can be identified which significantly differ in their entropic properties.
We are able to separate the contributions of the solvent and the chain configurational entropy explicitly to determine the influence on the unfolding pathways. The experimental results can be explained by 
a temperature dependent entropy for each conformation which is significantly dominated by the present number of hydrogen bonds with the solvent.\\ 
The paper is organized as follows. In the next section we present the numerical and experimental details. The results are presented in the fourth section. We conclude with a brief summary in the last section.
\section{Numerical Details}
We have performed our Molecular Dynamics simulations of the i-motif in explicit TIP3P solvent at 300 K by the GROMACS software package \cite{Gromacs} with the ffAmber03 force field \cite{Sorin2005}.
\begin{figure}[h!]
 \includegraphics[width=0.35\textwidth]{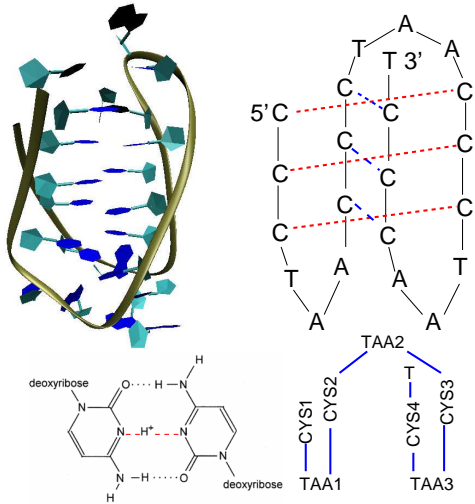}
 \caption{C-CH$^+$ pairing (bottom left) which is responsible for the formation of the DNA i-motif (top left) with the corresponding sequence (top right). The dashed lines represent the initially 
interacting nucleobases where C,T and A denote cytosine, adenine and thymine. A coarse grained model of the i-motif is shown in the right bottom.
}		
\label{fig1}
\end{figure}
The single DNA strand consists of 22 nucleic acid bases given by the sequence $5^{\prime}-CCC-[TAA-CCC]_3-T-3^{\prime}$ where $T$, $A$ and $C$ denote thymine, adenine and cytosine. 
We modeled this structure which is directly related to the sequence
used in \cite{Liu2003} 
by the PDB entry 1ELN \cite{PDB}.
The cubic simulation box with periodic boundary conditions has a dimension of $(5.41\times 5.41\times 5.41)$ nm filled with $5495$ TIP3P water molecules. 
The negative charge of $-22 e$ of the backbone has been compensated by $22$ sodium ions. We applied a Nose-Hoover thermostat to the system 
where all bonds have been constrained by the LINCS algorithm \cite{Gromacs}. Electrostatics have been calculated by the PME algorithm and the timestep was $2$ fs. 
After energy minimization by a steepest descend method, the initial
warm up phase of 1 ns has been performed by keeping the position of the DNA molecule restrained. 
For a detailed investigation of the unfolding mechanism we conducted five 300 K simulations each with 10 ns duration to calculate the average values.\\
The calculation of the free energy landscape has been performed by the metadynamics method presented in Ref.~\cite{Laio02}. The biased metadynamics simulations at 300 K have 
been conducted by the program plug-in PLUMED \cite{Plumed}. The Gaussian hills were set each 2 ps with a height of 0.25 kJ/mol and a width of 0.25 nm. The corresponding reaction coordinates for
the biased energy are the distance between nucleobase C1 and T22 and the distance between the center-of-mass for the combined nucleobases C1-T22 to the nucleobase A11. The final free energy landscapes
have been refined by histogram reweighting \cite{Kumar92} of 15 biased simulations of 10 ns at 300 K by the method introduced in Ref.~\cite{Smiatek}. The eigenvector free energy landscape has been calculated
by using a projection scheme \cite{Smiatek}.\\
To simplify our results for the kinetic investigations, we paired each three nucleobases in one group resulting in the 
sequence CYS1-TAA1-CYS2-TAA2-CYS3-TAA3-CYS4-T where CYS and TAA equals CCC, respectively TAA as shown as in Fig.~\ref{fig1}.\\
The calculation of the thermodynamic properties has been performed by keeping the position of each structure restrained for 100 ps.
\section{Experimental details}
The oligodeoxynucleotide was purchased from Sangon Co., Ltd. The sequence was identical to the simulations and it was dissolved in a final buffer with 50 mM MES and 50 mM NaCl. 
The buffer had a pH value of 8.0 and the 
concentration of DNA was 10 $\mu$M.
We used Circular Dichroism (CD) spectropolarimetry to investigate the structural behaviour. 
All CD measurements were recorded on Jasco Spectropolarimeter (J-810) equipped with a programmable temperature-control unit. The range of scanning wavelength was from 350 to 220 nm for
two different temperatures 298.15 K and 368.15 K.
\section{Results}
\subsection{Kinetic properties}
\begin{figure}[h!]
 \includegraphics[width=0.5\textwidth]{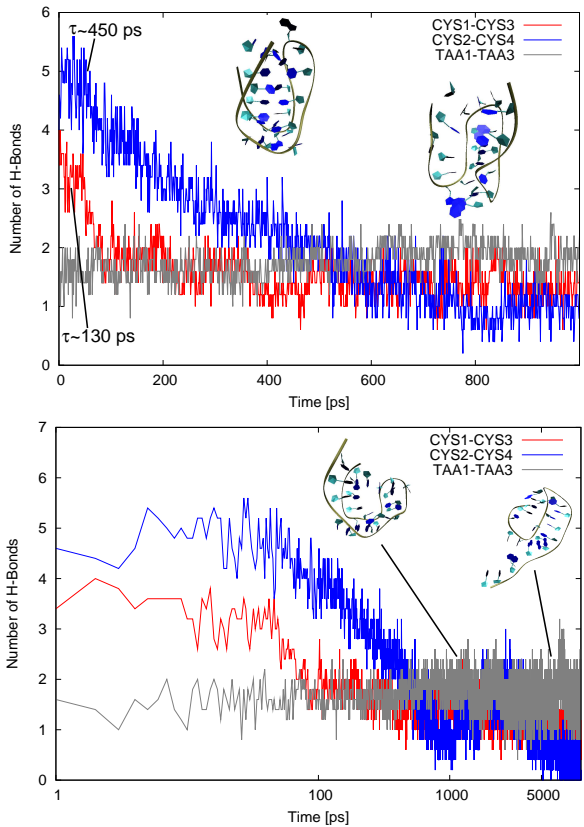}
 \caption{Number of hydrogen bonds between the groups CYS1-CYS3, CYS2-CYS4 and TAA1-TAA3 up to 1 ns (top) and 10 ns (bottom) averaged over 5 simulations. Typical configurations are shown as snapshots.  
  The initial vanishing of hydrogen bonds between the CYS2-CYS4 group can be roughly approximated by an exponential fit with $\tau\approx 450$ ps, respectively 130 ps for the CYS1-CYS3 groups.
}		
\label{fig2}
\end{figure}

We start this section by presenting the results for the unbiased simulations.
For this we calculated the number of hydrogen bonds between the CYS1-CYS3, CYS2-CYS4 and the TAA1 and TAA3 group (Fig.~\ref{fig1}) as a function of time shown in Fig.~\ref{fig2}. 
It has been reported \cite{Leroy2000,Liu2003} that the binding
between these groups is essential for the stability of the i-motif.
An initial decay of hydrogen bonds for the cytosine groups can be observed up to 1 ns. This vanishing can be roughly approximated by an exponential function $\exp(-t/\tau)$ where 
the constant can be fitted to $\tau\approx 450$ ps for the CYS2-CYS4 pair and $\tau\approx 130$ ps for the CYS1-CYS3 binding. The configurational snapshot at 600 ps shows a broadened i-motif. 
On longer timescales of 2-5 ns the $3^{\prime}$ end slightly opens. 
After 7 ns a stable configuration can be identified which is dominated by the open $3^{\prime}$ end with the CYS4-T groups and the $5^{\prime}$ end which is interacting with the central loop region.
The number of hydrogen bonds between the TAA1-TAA3 connection is nearly constant up to 10 ns. Thus the observed final structure can be best described
by a hairpin configuration \cite{Losick2004}. 
However regarding the short decay time of the initial hydrogen bonds, it becomes obvious that the starting structure in absence of hemi-protonated cytosines is very unstable in agreement to the results reported 
in Refs.~\cite{Gueron1993,Leroy2000,PDB,Liu2003,Mao2007}.\\
\begin{figure}[h!]
 \includegraphics[scale=0.35]{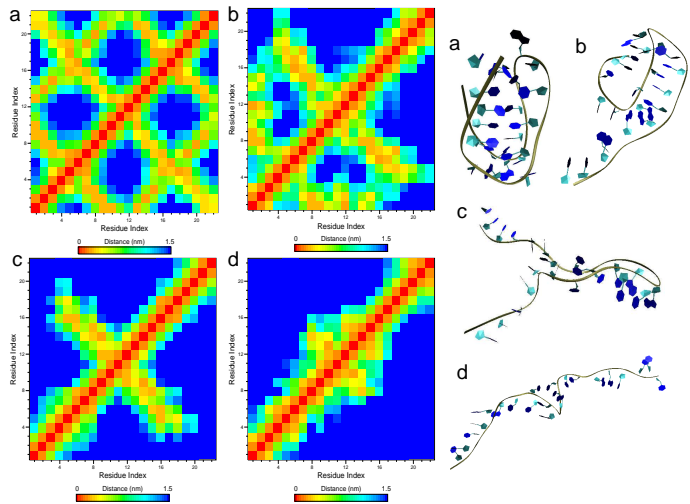}
 \caption{Distance matrices for the 22 nucleobases of the i-motif (left) and three other conformations. Panel a) shows the initial i-motif. Panel b) and c) present the results for two hairpin structures and Panel d) 
is related to a fully extended strand.
}		
\label{fig3}
\end{figure} 
To investigate the fully accessible phase space, we applied the metadynamics technique in which a history dependent biasing potential is applied to the molecule which helps to overcome energetic barriers. 
Details of the method can be found elsewhere \cite{Laio02,Laio08}.
We calculated the distance matrices for the nucleobases of the i-motif and further occurring conformations at later simulation times 
presented in Fig.~\ref{fig3}. It is obvious that the i-motif (Panel a) 
represents a well-defined 
structure with many local interactions even for long distances along the backbone. Two further structures (Panels b and c) differ in their nearest neighbor interactions. 
The structure of (b) is also shown in Fig.~\ref{fig2} after 7 ns and the structure of (c) is a fully planar hairpin structure which indicates cross-like interactions.
Local interactions for all nucleobases between C1-A12 can be observed in (b) whereas (c)
indicates interactions between the opposite sides of the strand. Hence these results are in good agreement to Fig.~\ref{fig2} where it is shown that the opening of the i-motif is 
also initialized by the $3^{\prime}$ end. 
In addition it has to be mentioned that we also have observed a fully extended configuration in our simulations (Panel d). This conformation is characterized by the neglect of the cross-like structure shown in Panel (c). 
It can be concluded that only local nearest neighbor interactions along the backbone are present with short distances between the nucleobases. Hence the vanishing of 
the i-motif can be easily rationalized in terms of distance matrices showing a decrease in the connectivity of the nucleobases for longer simulation times.\\
\subsection{Free energy landscapes}
The study of stable structures is achieved by the investigation of the free energy landscape. Generally, the calculation of free energy landscapes is a challenging task regarding problems 
like finite simulation time resulting in low sampling efficiencies for several regions as well as the neglect of structural transitions representing rare events \cite{Smiatek,Laio02}. 
Nevertheless, a lot of computational algorithms have been published all over the years to overcome these problems.\\
\begin{figure}[h!]
 \includegraphics[scale=0.3]{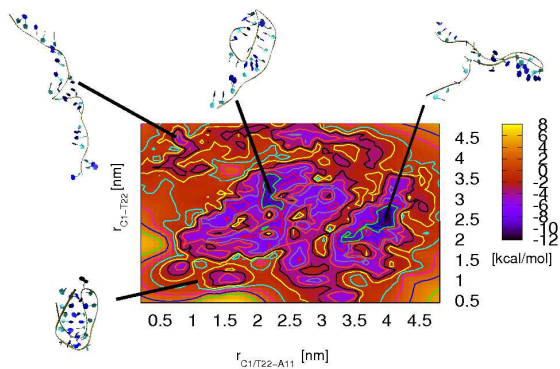}
 \caption{Free energy landscape for the distances between the center of mass for combined C1 and T22 and the distance to A11 $r_{C1/T22-A11}$ and the distance between C1 and T22 $r_{C1-T22}$ (top). The lines correspond 
to energy differences of 1.5 kcal/mol. 
}		
\label{fig4}
\end{figure}
Metadynamics allows to calculate the free energy landscape after
filling the free energy minima by an additional biasing energy in form of Gaussian hills. The Gaussians are set at each $\tau$ relaxation steps and 
are used to overcome energetic barriers such that the whole phase space is accessible. 
As collective variables we chose the distance between the 
center of masses of the A11 and the combined C1 and T22 nucleobases. As a second reaction coordinate we chose the distance between the center of masses for the C1 and the T22 group. 
We applied these observables due to the large variations in the distance matrices shown in Fig.~\ref{fig3}.
The results for the free energy landscape are presented in Fig.~\ref{fig4}.\\ 
Two large minima can be identified in a funnel-like landscape \cite{Karplus1998} with energy differences to the native structure around -8 kcal/mol. 
These results indicates them as very stable in contrast to the i-motif.
Regarding these conformations in detail, it becomes clear that these structures belong to the planar and partly planar structures 
shown in Fig.~\ref{fig3} (Panel b and c). 
The fully extended strand which was also observed in our biased simulations can be identified as energetically less
favourable.
By regarding the overlap of the two minima in each direction, it becomes clear that a separate 
calculation of the free energy for each coordinate would result in a significant decrease and error of the barriers and the minima \cite{Chandler02}. 
Hence only the usage of a two-dimensional representation leads to a sufficient estimate of the free energy differences.\\
We have also calculated the free energy landscape for the essential eigenvectors of the system \cite{Amadei93}. 
Eigenvectors have been shown as useful to capture the main concerted motion of the molecule \cite{Amadei93}. Mathematically 
they are closely related to principal component analysis. The main motion is described by the first eigenvectors which form the essential subspace whereas the fluctuating motion of the higher eigenvectors 
represents the remaining subspace. A detailed description can be found in Refs.~\cite{Smiatek,Amadei93}. 
To calculate the eigenvectors for the complete unfolding trajectories into the stretched structure we have performed five 500 K 
unbiased 
high temperature simulations each with 10 ns. 
The simulations have been combined to calculate an unbiased high temperature averaged eigenvector set. This set was used for the projection of the the free energy landscape onto the essential 
subspace of the first eigenvectors \cite{Amadei93,Smiatek} in the low temperature regime at 300 K. 
A recent publication \cite{SmiatekDNA2} has validated this technique.\\
The final two-dimensional energy landscape with the observed unfolding pathways is presented in Fig.~\ref{fig5}.
Two large minima with free energies around $\Delta F \approx -8$ kcal/mol belonging to conformations (b) and (c) can be identified. In contrast to them a local energetic minimum associated with a stretched structure (e)
given by a free energy difference $\Delta F \approx -5$ kcal/mol constitutes a metastable configuration.  
\begin{figure}
 \includegraphics[width=0.5\textwidth]{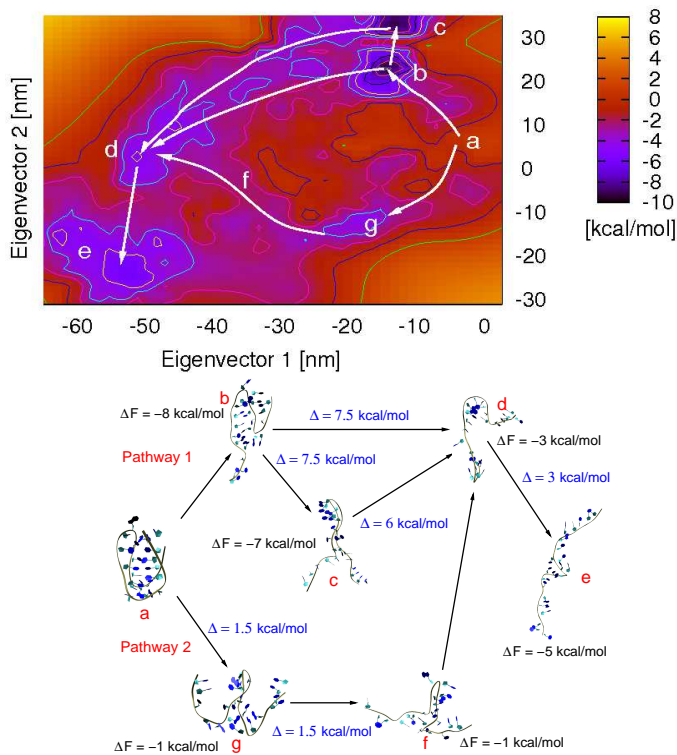}
 \caption{Free energy landscape for the eigenvectors 1 and 2 with the corresponding stable configurations (top) and unfolding pathways (bottom). 
The unfolding pathways with the corresponding free energy differences $\Delta F$ and free energy barriers $\Delta$ are shown in the bottom following
the notation in the top figure. 
  Transitions without given $\Delta$ values indicate the disappearance of barriers between the configurations.
}		
\label{fig5}
\end{figure}
These values are in good agreement to the results shown in Fig.~\ref{fig4}. By analyzing the biased metadynamics pathway and corresponding high temperature unfolding simulations \cite{SmiatekDNA2}, 
we are able to discover the main unfolding pathways as presented in the bottom of Fig.~\ref{fig5}.\\ 
Pathway 1 follows the spontaneous unfolding into hairpin structure (b) with a free energy difference of $-8$ kcal/mol without
notable energy barriers. From structure (b) one option is to unfold into structure (c) with a barrier of $\Delta=7.5$ kcal/mol or directly transform into structure (d) with identical values for the barrier height. 
Having unfolded into
structure (d), the s-shape form can transform into a fully extended strand (e) with a free energy difference of $-5$ kcal/mol by overcoming a free energy barrier of 3 kcal/mol.
It has to be mentioned that structural transformations into (d) and (e) are largely hindered due to energetic arguments
such that 
it can be concluded that the hairpin structures represent the stable conformations at 300 K.\\
The other pathway is realized by a torsional motion into structure (g) which is energetically separated by 1.5 kcal/mol from the i-motif.
The transition from structure (g) into (f) is hindered by an energy barrier of roughly $\Delta \approx 1.5$ kcal/mol until it freely unfolds into 
structure (d) to join the first pathway into the fully extended strand. Combining all results, it finally can be concluded that the global minima
given by the hairpin configurations are energetically more favourable than other
structures. Additionally it has to be mentioned that these conformations are stabilized due to large energy barriers which are in the range of the free energy differences.\\   
The one-dimensional representation for each eigenvector calculated by a projection scheme \cite{Smiatek} is presented on the left side of Fig.~\ref{fig6} whereas the corresponding 
concerted motion is shown on the right side. 
It can be seen that eigenvector 1 describes the variation of the end-to-end distance by a stretching motion whereas eigenvector
2 mainly captures the relaxation from the i-motif to the planar structure.\\ 
The deepest minimum 
along eigenvector 1 can be found for the hairpin configurations where the values are in good agreement to the results derived in Fig.~\ref{fig5}. It is obvious that the averaged 
barrier height between the hairpin structures and the stretched
structures is above 6 kcal/mol. For a detailed investigation of the two hairpin structures we have calculated the average values for eigenvector 2 along a window 
of eigenvector 1 between -20 nm and -10 nm resulting in an averaged calculated barrier height of around 5 kcal/mol. 
Hence it can be concluded, that after regarding all results the hairpin structures represent the most energetically stable configurations.
\begin{figure}
 \includegraphics[width=0.5\textwidth]{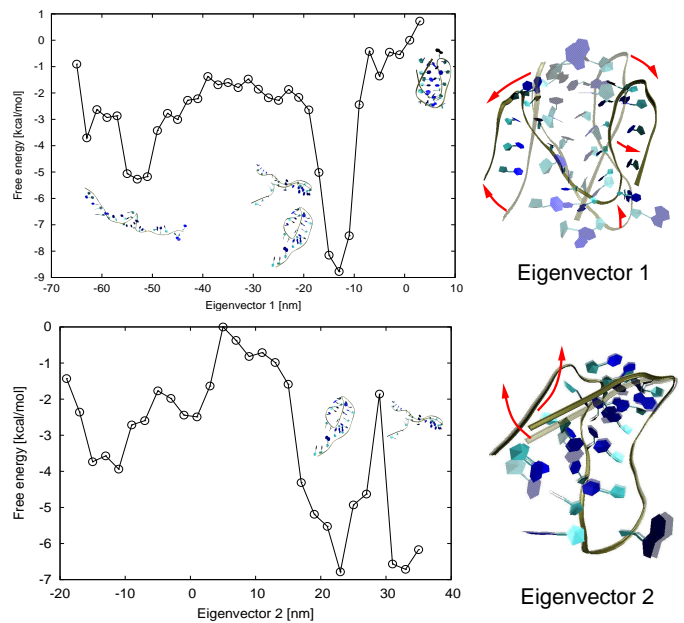}
 \caption{Free energy landscape for the one-dimensional representation along eigenvector 1 (top left) and eigenvector 2 (bottom left) with the corresponding conformations. The values for eigenvector 2 have been computed
along the values -20 nm to -10 nm of eigenvector 1 to compute the average barrier between the hairpin structures. The concerted motion of the eigenvectors 1 and 2 is shown in the right as indicated by the red arrows.
}		
\label{fig6}
\end{figure} 
\subsection{Thermodynamic properties}
For a closer view on the unfolding mechanisms,
we have calculated the total energy and the temperature entropy contribution to the free energy in Tab.~\ref{tab1} for each configuration shown in Fig.~\ref{fig5} relative to the i-motif. 
The total entropy is given by the 
relation $\Delta S=(\Delta U - \Delta F)/T$ where $\Delta F$ and $\Delta U$ denote the free and the total energy differences whereas $T$ represents the systems temperature. 
It becomes clear that the most energetically favourable 
conformations are given by structures (d), (f) and (g) which are representing the second pathway. 
However, the large deficit of entropy compared to the other structures 
indicates these conformations and in general the second pathway as unfavorable in agreement to the results shown in 
Fig.~\ref{fig5}.
Compared to the i-motif, it is surprising that all other structures are entropically less favourable. This can be related to structural changes in the surrounding water shell as we will show in the following.\\ 
For this we have calculated the entropic contributions by the relation $\Delta S = \Delta S_e + \Delta S_c$ separately where $\Delta S_e$ represents the change in the environmental 
entropy and $\Delta S_c$ the change in the configurational entropy calculated by a quasi-harmonic approach \cite{Karplus2001}. The results are shown in Tab.~\ref{tab2}.
Both pathways are comparable in their configurational entropy gain. However, a significant
contribution to the total entropy arises from local environment changes $\Delta S_e$. All values are negative, leading 
to the conclusion that the water around the initial i-motif has a larger entropy. The largest deviations occur for structures (d), (f) and (g) belonging to the second pathway.
This also supports the fact that the second pathway is entropically less preferable.\\
The variation of $\Delta S_e$ and the total energy difference $\Delta U$ can be rationalized
by regarding the values for the number of hydrogen bonds relative to the i-motif $\Delta n_H$. 
Regarding the values of Tab.~\ref{tab2} it becomes clear that a large number of hydrogen bonds between the solvent and the DNA leads to a significant deficit of entropy due to local ordering and to a 
negative increase of the total energy.
It is significant that the hairpin structures (b) and (c) show the smallest increase of $\Delta n_H$. This can be explained by the fact that most of the nucleobases belonging to these conformations 
are not accessible to water molecules leading to an avoidance of additional hydrogen bonds. Structures like (d), (g) or (f) are more hydrated as it can be seen by the increased 
number of hydrogen bonds which results in larger entropy
deficits of the environmental entropy. Thus the main unfolding pathway and the stability of the hairpin conformations can be mainly explained by entropically more preferable structures due to a 
diminished number of hydrogen bonds compared to other configurations.\\
\begin{table}
\caption{Total energy $\Delta U$ and temperature-entropy contribution $T\Delta S$ to the free energy difference for the different configurations shown in Fig.~\ref{fig5} relative to the 
i-motif structure.}
\label{tab1}       
\begin{tabular}{cccc}
\hline\noalign{\smallskip}
& $\Delta U$ [kcal/mol] & $T\Delta S$ [kcal/mol] & $\Delta F$ [kcal/mol] \\
\noalign{\smallskip}\hline\noalign{\smallskip}
b & $-59$ & $-51$ & -8\\
\noalign{\smallskip}\hline\noalign{\smallskip}
c & $-21$ & $-14$ & -7 \\
\noalign{\smallskip}\hline\noalign{\smallskip}
d & $-127$ & $-124$ & -3\\
\noalign{\smallskip}\hline\noalign{\smallskip}
e & $-62$ & $-57$ & -5\\
\noalign{\smallskip}\hline\noalign{\smallskip}
f &  $-140$ & $-139$ & -1\\
\noalign{\smallskip}\hline\noalign{\smallskip}
g &  $-177$ & $-176$ & -1\\
\noalign{\smallskip}\hline
\end{tabular}
\end{table}
\begin{table}
\caption{Total entropy $\Delta S$, configurational entropy $\Delta S_c$, environment entropy $\Delta S_e$ and number of hydrogen bonds $\Delta n_H$ for the different configurations shown in Fig.~\ref{fig5} relative to the 
i-motif structure}
\label{tab2}       
\begin{tabular}{ccccc}
\hline\noalign{\smallskip}
& $\Delta S$ [kcal/K mol] & $\Delta S_c$ [kcal/K mol]  & $\Delta S_e$ [kcal/K mol] & $\Delta n_H$\\
\noalign{\smallskip}\hline\noalign{\smallskip}
b & $-0.17$ & $0.35$ & $-0.52$ & 11 \\
\noalign{\smallskip}\hline\noalign{\smallskip}
c & $-0.05$ & $0.34$ & $-0.39$ & 1 \\
\noalign{\smallskip}\hline\noalign{\smallskip}
d & $-0.41$ & $0.34$ & $-0.75$ & 20\\
\noalign{\smallskip}\hline\noalign{\smallskip}
e & $-0.19$ & $0.35$ & $-0.54$ & 31\\
\noalign{\smallskip}\hline\noalign{\smallskip}
f &  $-0.46$ & $0.34$ & $-0.80$ & 25\\
\noalign{\smallskip}\hline\noalign{\smallskip}
g &  $-0.59$ & $0.33$ & $-0.92$ & 37\\
\noalign{\smallskip}\hline
\end{tabular}
\end{table}
To understand the drastic variation of the free energy landscape for higher temperatures which has been reported in Ref.~\cite{SmiatekDNA2}, we have analyzed the thermodynamic properties at 500 K.
It has been shown \cite{SmiatekDNA2}, that at these temperatures the hairpin structures are only metastable configurations and an extended structures similar to (e) represents the global equilibrium conformation.
To study these properties in detail we repeated the thermodynamic calculations with identical structures for a temperature of 500 K.
The values for the total energy, the free energy differences and the temperature-entropy configuration at 500 K are shown in Tab.~\ref{tab3}.
\begin{table}
\caption{Total energy $\Delta U$ and temperature-entropy contribution $T\Delta S$ to the free energy difference for the different configurations shown in Fig.~\ref{fig5} relative to the 
i-motif structure at 500 K.}
\label{tab3}       
\begin{tabular}{cccc}
\hline\noalign{\smallskip}
& $\Delta U$ [kcal/mol] & $T\Delta S$ [kcal/mol] & $\Delta F$ [kcal/mol] \\
\noalign{\smallskip}\hline\noalign{\smallskip}
b & $-31$ & $-29$ & -2\\
\noalign{\smallskip}\hline\noalign{\smallskip}
c & $-7$ & $-4$ & -3 \\
\noalign{\smallskip}\hline\noalign{\smallskip}
d & $-48$ & $-45$ & -3\\
\noalign{\smallskip}\hline\noalign{\smallskip}
e & $-10$ & $-7$ & -3\\
\noalign{\smallskip}\hline\noalign{\smallskip}
f &  $-47$ & $-45$ & -2\\
\noalign{\smallskip}\hline\noalign{\smallskip}
g &  $-108$ & $-106$ & -2\\
\noalign{\smallskip}\hline
\end{tabular}
\end{table}
It comes out that nearly all values compared to Tab.~\ref{tab1} are smaller. This can be explained by a decreased number of hydrogen bonds shown in Tab.~\ref{tab4} 
which interact with the i-motif due to an increased thermal energy resulting
in smaller Bjerrum lengths \cite{SmiatekDNA2}. Additionally it can be assumed that the dependence of the entropy on the number of hydrogen bonds is less pronounced for 500 K compared to 300 K. 
Hence the values for the total energy and the temperature-entropy contribution are lowered which leads to a change of the free energy landscape as it can be seen by the 
values of $\Delta F$ in Tab.~\ref{tab3} compared to Tab.~\ref{tab1}.\\ 
\begin{table}
\caption{Total entropy $\Delta S$, configurational entropy $\Delta S_c$, environment entropy $\Delta S_e$ and number of hydrogen bonds $\Delta n_H$ for the different configurations shown in Fig.~\ref{fig5} relative to the 
i-motif structure at 500 K}
\label{tab4}       
\begin{tabular}{ccccc}
\hline\noalign{\smallskip}
& $\Delta S$ [kcal/K mol] & $\Delta S_c$ [kcal/K mol]  & $\Delta S_e$ [kcal/K mol] & $\Delta n_H$\\
\noalign{\smallskip}\hline\noalign{\smallskip}
b & $-0.05$ & $0.44$ & $-0.49$ & 11 \\
\noalign{\smallskip}\hline\noalign{\smallskip}
c & $-0.01$ & $0.43$ & $-0.44$ & 1 \\
\noalign{\smallskip}\hline\noalign{\smallskip}
d & $-0.09$ & $0.44$ & $-0.53$ & 20\\
\noalign{\smallskip}\hline\noalign{\smallskip}
e & $-0.01$ & $0.45$ & $-0.46$ & 23\\
\noalign{\smallskip}\hline\noalign{\smallskip}
f &  $-0.09$ & $0.43$ & $-0.52$ & 17\\
\noalign{\smallskip}\hline\noalign{\smallskip}
g &  $-0.21$ & $0.43$ & $-0.61$ & 26\\
\noalign{\smallskip}\hline
\end{tabular}
\end{table}
To investigate this in more detail we have calculated the entropic contributions  for each configuration at 500 K. The results are presented in Tab.~\ref{tab4}. 
All entropy values are smaller except the configurational entropy difference $\Delta S_c$ coming from thermal activation of further degrees of freedom at elaborated temperatures.
Comparing $\Delta n_H$ and $\Delta S_e$ indicates that the differences between the configuration are less pronounced in contrast to lower temperatures (Tab.~\ref{tab2}) as it was discussed above. 
This can be related to a weaker dependence of $\Delta S_e$ on $\Delta n_H$ and, second on a general decrease of $\Delta n_H$ for certain strucutures.  
Hence the general change of the entropy for higher temperatures as mentioned above can be mainly understood by a variation of $\Delta S_e$ alone. It can be seen that the dependence of hydrogen bonds on the 
environmental entropy at higher temperatures is less important.
In addition due to nearly identical entropy values for each structure, 
the contributions of 
the total energy become more important which results in the small variations of the free energy differences shown in Tab.~\ref{tab3}. Hence our results have shown that the variation of the free energy landscape at higher 
temperatures is caused by a temperature-dependent entropy which is largely influenced by local hydrogen bonds with the solvent.\\
To study the properties and the characteristics of $\Delta n_H$ for different situations and temperatures, we have calculated the ratio of hydrogen bonds compared to the i-motif for backbone and non-backbone hydrogen bonds 
individually.
\begin{figure}
 \includegraphics[width=0.5\textwidth]{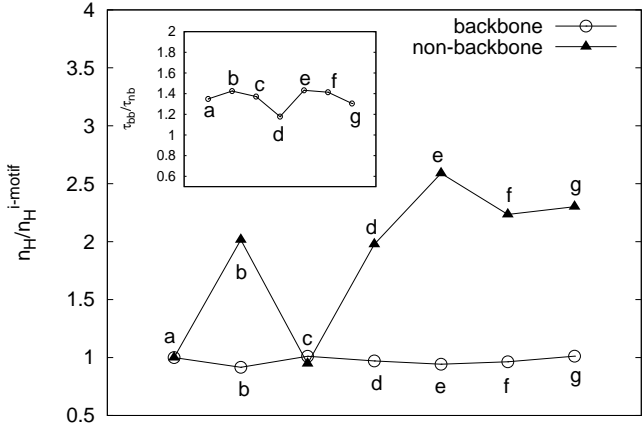}
 \caption{Ratio of the backbone and non-backbone hydrogen bonds $n_H$ compared to the i-motif $n_H^{i-motif}$ for each configuration at 300 K. {\bf Inset:} Ratio of the lifetimes of the backbone hydrogen bonds 
$\tau_{bb}$ and the non-backbone hydrogen bonds $\tau_{nb}$ for each configuration at 300 K.  The lines are only for the eyes.
}		
\label{fig7}
\end{figure}
The results are presented in Fig.~\ref{fig7} for 300 K. It is obvious that the backbone ratio is nearly constant for each structure. A significant increase of non-backbone hydrogen bonds coming from 
nucleobase water interactions can be observed for 
all structures except (c). In addition, several structures show an increase of the accessible hydrophilic surface area of 7 (b), 11 (d) and 6 (f,g) nm$^2$ compared to the i-motif which also supports the results 
shown in Fig.~\ref{fig7}. 
Additionally it can be assumed that the hydrogen bonds of the backbone with the water are energetically more stable due to larger electrostatic interactions. This becomes obvious 
by regarding the lifetimes of the hydrogen bonds for the backbone $\tau_{bb}$ and for non-backbone atoms $\tau_{nb}$. The ratio $\tau_{bb}/\tau_{nb}$ is shown in the inset of Fig.~\ref{fig7}. 
The reason for this behaviour can be explained by higher charged backbone atoms in comparison to lower charged nucleobase atoms which lead to energetically 
more stable bonds.\\
As it was discussed above for 300 K, the variation of the environmental entropy for each structure is mainly induced 
by additionally accessible nucleobase atoms forming hydrogen bonds with the solvent. 
Their properties at higher temperatures are presented in Fig.~\ref{fig8}
where the ratio of the number of hydrogen bonds at 500 K and 300 K $n_{H,500K}/n_{H,300K}$ is shown.
It becomes clear that lower ratios are given for non-backbone hydrogen bonds indicating weaker interactions. By comparison with the results for $\Delta n_H$ shown in Tabs.~\ref{tab2} and \ref{tab4} 
it can be seen that the deficit of $\Delta n_H$ at higher temperatures is mainly given by a decrease of non-backbone hydrogen bonds resulting in the variation of the environmental entropy.\\
By combining all results, it becomes obvious that the environmental entropy for the unfolded conformations is significantly lowered for certain conformations compared to the i-motif due to 
additionally accessible non-backbone hydrogen bonds. 
These bonds are energetically less stable compared to the 
backbone hydrogen and are therefore diminished at higher temperatures. This fact leads to smaller variations of the environmental entropy compared to lower temperatures which results in different free energy 
landscapes as it has been observed in Ref. \cite{SmiatekDNA2}. 
\begin{figure}
 \includegraphics[width=0.5\textwidth]{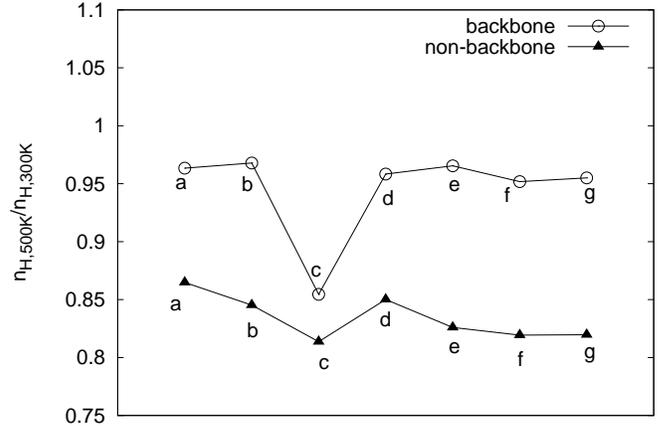}
 \caption{Ratio of the number of hydrogen bonds $n_{H,500K}/n_{H,300K}$ at 500 K and 300 K for non-backbone and backbone atoms. The lines are only for the eyes.
}		
\label{fig8}
\end{figure}
\subsection{Experimental results}
\begin{figure}
 \includegraphics[width=0.5\textwidth]{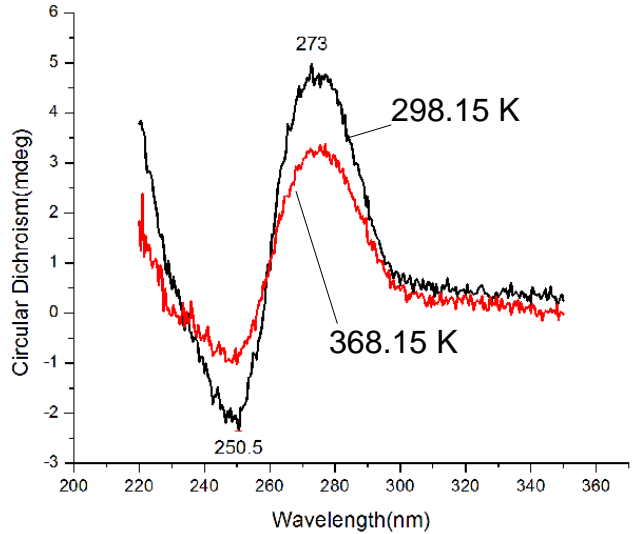}
 \caption{CD spectra at 298.15 K (black) and 368.15 K at pH value 8.0. The small numbers denote the positions of the minima and maxima.
}		
\label{fig9}
\end{figure}

The stability of the hairpin structures at 300 K and the appearance of the extended structures at higher temperatures is also supported by regarding the results of circular dichroism (CD) spectropolarimetry. 
The CD spectra at two different temperatures 298.15 K and 368.15 K are presented in Fig.~\ref{fig9}.
It has been reported in \cite{Brahmachari01,Sugiyama06} that the DNA i-motif has a maximum at 285 nm, a negative minimum at 260 nm and a crossover at 270 nm. Hence the results shown in Fig.~\ref{fig9} display an 
absence of the i-motif which is indicated by a shifted maximum at 273 nm, a minimum at 250.5 nm and a crossover at 257 nm. In pioneering studies \cite{Shapiro70} it has been discussed that a maximum at 270 nm
corresponds to a random-coil structure which could be also brought into accordance to structures (b) and (c) of Fig.~\ref{fig5}. Related to this, a further study has also assumed the existence of stable hairpin structures 
at pH values of 7.2 based on their 
experimental CD results \cite{Brahmachari01}.\\ 
It was in general validated \cite{Marky02,Lewis07} that the intensity of the minima is given by the number of connected base pairs and intramolecular interaction  energies. 
As it can be seen in Fig.~\ref{fig9}, the magnitude of the minimum and the maximum depends on the applied conditions which indicates a decrease of intramolecular interactions for higher temperatures. 
This is in agreement to a 
recent publication \cite{SmiatekDNA2} and to the results of Fig.~\ref{fig3} and the discussion above where it has been shown that the global stable conformation at higher temperatures is given by the fully extended strand 
where intramolecular interaction energies are 
largely negligible. 
The reason
for the change of the global minima can be explained by a temperature dependent entropy which favours the expanded conformation as it 
has been validated by the previous results.\\
Base stacking energies also have been reported \cite{Marky02} as responsible for larger magnitudes at 255 nm   
and specifically for mismatched hairpin configurations. 
Additionally it was found that homomismatches of C-C pairing are energetically more favourable than other hetero mismatches explaining this unusual stability \cite{Marky02}.\\ 
The observed CD spectrum can be brought into accordance with a hairpin B-DNA structure \cite{Benight89} and it can be also assumed 
that the T-A pairing, which has been validated by the results of Fig.~\ref{fig2} and Fig.~\ref{fig3} is mainly responsible for the stability of these special structures.
Furthermore it has to be noticed that the i-motif is absent due to negative CD spectra minima around 297 nm which belong to deprotonated cytosines \cite{Chowdhury04}.
Hence it can be concluded, that at basic conditions the DNA i-motif is absent and hairpin structures represent the global minima. Our experiments have shown that these configurations transform into a fully extended 
strand at elaborated temperatures in good agreement to the numerical results.
\section{Summary and conclusion}
We have simulated the unfolding of an unprotonated DNA i-motif at 300 K via Molecular Dynamics simulations in explicit solvent. 
Our results indicate the planar and partly planar configurations as the most stable structure in absence of protonated cytosines. These structures are realized by an individual opening of each end of the strand.
The vanishing of the intramolecular hydrogen bonds which are responsible for the formation of the 
DNA i-motif occurs on a time scale which is significantly shorter than 10 ns. This validates that the deprotonated i-motif is not stable at room temperature in agreement to earlier published experimental 
results \cite{Leroy2000,Tan2005}. A stable partly planar configuration appears on a timescale of 7 ns.\\
The calculation of the essential dynamics in combination with the corresponding free energy landscape allows to determine the unfolding pathways.
Two main unfolding mechanisms can be identified where the preferential one is entropically more favourable leading to lower free energies.
We have shown that large contributions to the entropy values arise from local water entropy deficits which are caused by hydrogen bonds. 
The nature of these bonds and their importance for a temperature-dependent entropy has
been investigated in detail. 
Hence we have shown that the prior unfolding process is mainly determined by entropic contributions dominated by lower DNA-solvent ordering interactions.\\
The numerical results are in good agreement to experimental CD spectra. 
It can be shown that the temperature dependence of the experimentally observed results can be explained by a vanishing of hairpin structures at higher temperatures due to a variation of the free energy landscape.
By analyzing the spectral data, it becomes 
obvious that base stacking energies largely dominate intramolecular interactions at lower temperatures which also validates the presence of hairpin structures \cite{Marky02}.\\ 
Thus our results allow to improve the usage of i-motifs in nanomachines due to the estimation of the unfolding pathway and its stable configurations. 
As it has been discussed above, we have validated that a full unfolding into a stretched structure is not energetically favourable and can be only realized by thermal or energetic activation. 
This sheds a new light on the importance of the unfolding 
pathway and its stable configurations.
It was assumed in Ref.~\cite{Liu2003,Zhang2010} that the unfolding process of an i-motif leads to a fully extended conformation which is in contrast to our results. 
Although it can be concluded that the presence of a large number of grafted i-motifs may lead to different unfolding pathways due to interchain interactions,
our results allow to optimize the improvement for low grafting densities which are needed for the fabrication of nanocontainers \cite{Mao2007}. 
We currently have undertaken attempts to investigate these properties in more detail.\\ 
However, even in a biological context our results can be adopted for a further study of 
possible i-motif configurations in the human cell. As it has been discussed in the introduction, this may be important for a pharmaceutical point of view due to the fact 
that the upstream of the i-motif  
along the insuline gene is related to its conformation and therefore facilitates 
transcription \cite{Gupta1997}.
Hence we hope that our results allow to achieve 
a deeper insight into the mechanisms of i-motif formation in vivo and vitro. 
\section{Acknowledgements}
We thank the Deutsche Forschungsgemeinschaft (DFG) through the transregional collaborative research center
TRR 61 for financial funding. 

\end{document}